# Mapping Temporal Trends of Parent-Child Migration from Population-Scale Family Trees

**Caglar Koylu and Alice Kasakoff**

Geographical and Sustainability Sciences, University of Iowa, Iowa City, USA
Geography, University of South Carolina, Columbia, USA
* caglar-koylu@uiowa.edu

## Introduction

User-generated family trees are invaluable for constructing population-scale family networks and studying population dynamics over many generations and far into the past (Han et al., 2017; Kaplanis et al., 2018; Kasakoff, 2019). Family trees contain information on individuals such as birth and death places and years, and kinship ties, e.g., parent-child, spouse, and sibling relationships. Such information about individuals in family trees makes it possible to extract migration networks over time. Despite the recent advances (Andrienko et al., 2017; von Landesberger et al., 2016), existing spatial and temporal abstraction techniques for time-variant flow data have limitations due to the lack of knowledge on the effect of temporal partitioning on flow patterns (Çöltekin et al., 2011).

In this study, we extracted state-to-state migration patterns over a period of 150 years between 1776 and 1926 from a cleaned, geocoded and connected family trees from Rootsweb.com (Koylu et al., 2020). We used birthplaces and birthyears of parents and children to extract intergenerational migration flows between states. To reveal the temporal trends of migration patterns, we evaluated three temporal partitioning strategies: (1) predefined periods in American history, (2) overlapping time periods with fixed length, and (3) time periods with variable length, which have approximately equal volume of moves per time period. To account for the effect of geographic proximity and flow volumes in migration flows, we transformed the raw flows into modularity flows (Newman, 2006) using a double-constrained a gravity model (Roy & Thill, 2004). Our preliminary results revealed longitudinal population mobility in the U.S. on such a large spatial and temporal scale.

## Method

We extracted migration patterns by taking the parents' birth state or territory as the origin and the child's birth state or territory as the destination. There is evidence that larger families moved farther and more often than smaller families both to secure land for their sons and because having several sons reduced the cost of labor to clear land near the frontier. Moves were frequently undertaken when the father was in his 40's or 50's to maximize family labor (Adams & Kasakoff, 1984). To reduce the bias of large families, we counted the four gender categories of parent-child relations once for those instances in which a parent had multiple children with the same birth state and gender. If the same sex children were born in the same state, mother-child and father-child relations were counted only once. We employed three temporal partitioning strategies to capture flow patterns and their change over time.

Although our data go back to early 17th century, we used 1776, The Revolutionary War as the starting point of the time range of our data to alleviate the challenges of mapping and analysis of the changing borders of states and territories. We further rounded the ending of the time range to 1926 to include a 150-year period. First, we partitioned the time into four periods based on breakpoints in American history: 1776-1820, 1820-1862, 1862-1890 and 1890-1926. The periods start with the Revolutionary War (1776). The first break (1820) is the mid-point between the Revolutionary War and the Civil War. The Civil War (1862) was the next break point that was followed by 1890 because the Census Bureau announced that the frontier had ended at that date. Second, we partitioned the time using fixed length or equal intervals, i.e., 15, 20, 25..., 60 years. This generated time periods such as 1776-1786, 1786-1796…, 1916-1926 for the 20-year fixed periods. Third, we extracted time periods with variable length such that each time period has approximately equal volume of flows.

We conducted a modularity-based evaluation of the three partitioning strategies. Modularity measure is calculated by the difference between the actual flows and the expected flows obtained from a null model (Newman, 2006). The modularity between a pair of states (i and j) is calculated by:

$$\text{Modularity (i, j)} = \text{Observed Flows } (F_{ij}) - \text{Expected Flows } (E_{ij})$$

where $F_{ij}$ is the number of observed flows and $E_{ij}$ is the expected number of flows from state i to state j. The flow between a pair of states is above expectation if modularity value is positive and below expectation if the value is negative. We calculated the expected flows using a double-constrained gravity model that constrains both origins and destinations and forces: (1) the sum of expected flows from an origin is equal to the observed, and (2) the sum of expected flows to a destination is equal to the observed volume of flows to that destination.

$$E_{ij} = A_i * O_i * B_j * D_j * D_{ij}^{-beta}$$

where $O_i$ and $D_i$ are the total of out-flows and in-flows of state i, $A_i$ and $B_i$ are the balance factors that are calculated by the following iteration. Each state has a different set of parameters. The distance decay function is square and uniform for all states.

$$A_i = 1 / \sum_{i=0}^{n} \sum_{j=0}^{n} (B_j D_j * D_{ij}^{beta})$$

$$B_i = 1 / \sum_{i=0}^{n} \sum_{j=0}^{n} (A_j O_j * D_{ij}^{beta})$$

**Results**

The data set we produced in our original study (Koylu et al., 2020) includes about 80 million individuals. In this study, we extracted parent-child relations using all of these 80 million records, but only included the relations that the migration year (birth year of the child -2) were between 1776 and 1926. Using the strategy for counting the parent-child relationship once for a parent who had multiple children with the same birth state and sex, we extracted about 41.5 million parent-child pairs with existing birth years. We were able to locate both the parent and child location for about 34.4 million pairs (~83%) which included foreign-born parents or children. Out of 34.4 million parent-child relations, 34% was between the states, 25% was within the U.S., and about 41% were to or from a foreign country.

Figure 1 illustrates the total modularity produced by each partitioning method and number of periods. The number of periods can be defined for the equal volume partitioning, whereas fixed-length partitioning creates periods with varying length depending on the fixed-length of each period. We use 4 periods to be able to compare the two methods with the historical partitioning. Equal volume partition produced the highest modularity than both the fixed-length and historical partitions with 4 periods. However, historical periods have a very close value. Also, fixed-length partitioning with 5, 6, 8 and 10 periods produced slightly larger modularity than the equal-volume periods with the same number of periods. In this paper, we analyze the historical periods to improve the interpretation of the patterns.

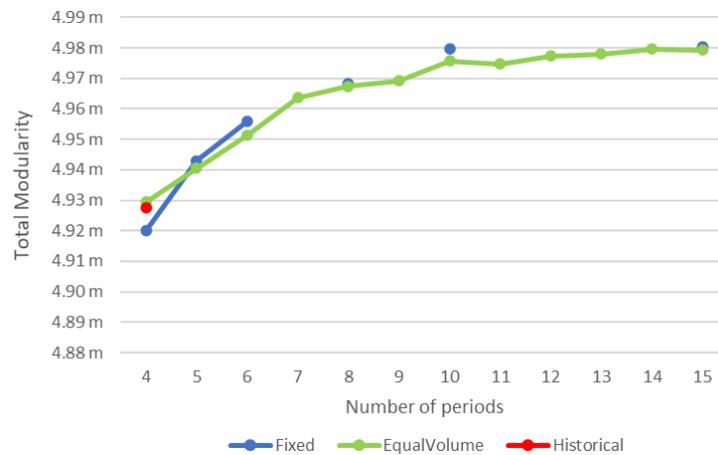

Figure 1: Total modularity of the three partitioning strategies using equal number of partitions: historical periods, fixed-length periods and equal volume periods.

Because the state and territorial boundaries in the U.S. evolved substantially during the study period, in this paper, we only illustrate flows in a 150-year period between 1776 and 1926 (Figure 2). Flows are curvy at the origin and straight at the destination end with a half-arrowhead to enhance the readability of the flow lines (Koylu & Guo, 2017). While the choropleth map illustrates netflow ratio of parent-child relations (see equation below), circle sizes are proportional to the number of relations within states, and flows illustrate the modularity values for between state relations.

$$Netflow\ Ratio_i = (Inflow_i - Outflow_i)/(Inflow_i + Outflow_i)$$

The predominant flows at all periods were East to West. These divided the US east of the Mississippi into horizontal bands which resulted in cultural and dialectical regions that still exist today. The first period (1776 – 1820) highlights Massachusetts and Virginia as centers. Interestingly, these are the two classical states scholars in Colonial History have focused on, epitomizing the North and the South. But here, the two states are in the period after the Colonial period with the most rooted populations. In the second period (1820 – 1862), Massachusetts no longer had a lot of within state pairs but New York, Pennsylvania, North Carolina joined Virginia which had a lot, and a second wave emerged in Ohio, Kentucky, and Tennessee. In the third period (1862 – 1890), the states experiencing net out-migration expanded to the border of Mississippi whereas the territories in the West experienced high net in-migration. The fourth period (1890 – 1926) is similar to the third with more flows to the West. The last period is the period of urbanization but the flows in the final map are still largely East to West. But interstate migration had lessened, and the difference between Eastern losses and Western gains became less extreme.

**Conclusion and Future Work**

We presented a preliminary analysis of migration through parent-child relations for individuals that span across several centuries. Since the U.S. population was so mobile during the 19th century at least half the segment of the population whose parents had been born in the U.S. lived in a different state from where their parents had been born. There are a number of future directions for our preliminary study. First, our estimation of migration two years before the childbirth is an initial choice but requires further evaluation. The dating of moves during the long period from a parent's birth to the births of their children makes it difficult to align our results with more precise measures. Some states were both sending people West and receiving from the East in each period. Likely the moves in preceded the moves out but the breadth of the historical periods made it impossible to see this effect. Alternative to parent-child migration, we plan to use the child-ladder approach (Lathrop, 1948) to extract migration using changes in birthplaces of consecutive siblings in a family. Second, we used the set historical periods, fixed and variable length time periods to extract migration patterns over time. Every time partition is meaningful. It is interesting, however, that the broad historical periods used by historians performs comparable to other partitioning methods. In a way, the importance of key events such as the Civil War and the closing of the frontier, has been validated through our comparison with other ways of partitioning time. Third, in addition to using modularity to maximize the flow patterns within time periods, we plan to systematically evaluate the changes in flow volumes and structures using temporal natural breaks, persistence measures (Bazzi et al., 2016; Pamfil et al., 2019; Roth et al., 2011), and the goodness of absolute deviations from the median (Slocum et al., 2009). Fourth, we plan to analyze how migration affected the network of family connectedness, particularly the density of kin in space and probability of having kin nearby to provide social and economic support (Koylu et al., 2014). Ravenstein observed in the 1880 British census (Ravenstein, 1885) that women moved more frequently than men but at shorter distances. Fifth, we plan to disaggregate flows by gender into mother-daughter, mother-son, father-daughter and father-son relations and study gender effects on migration over time.

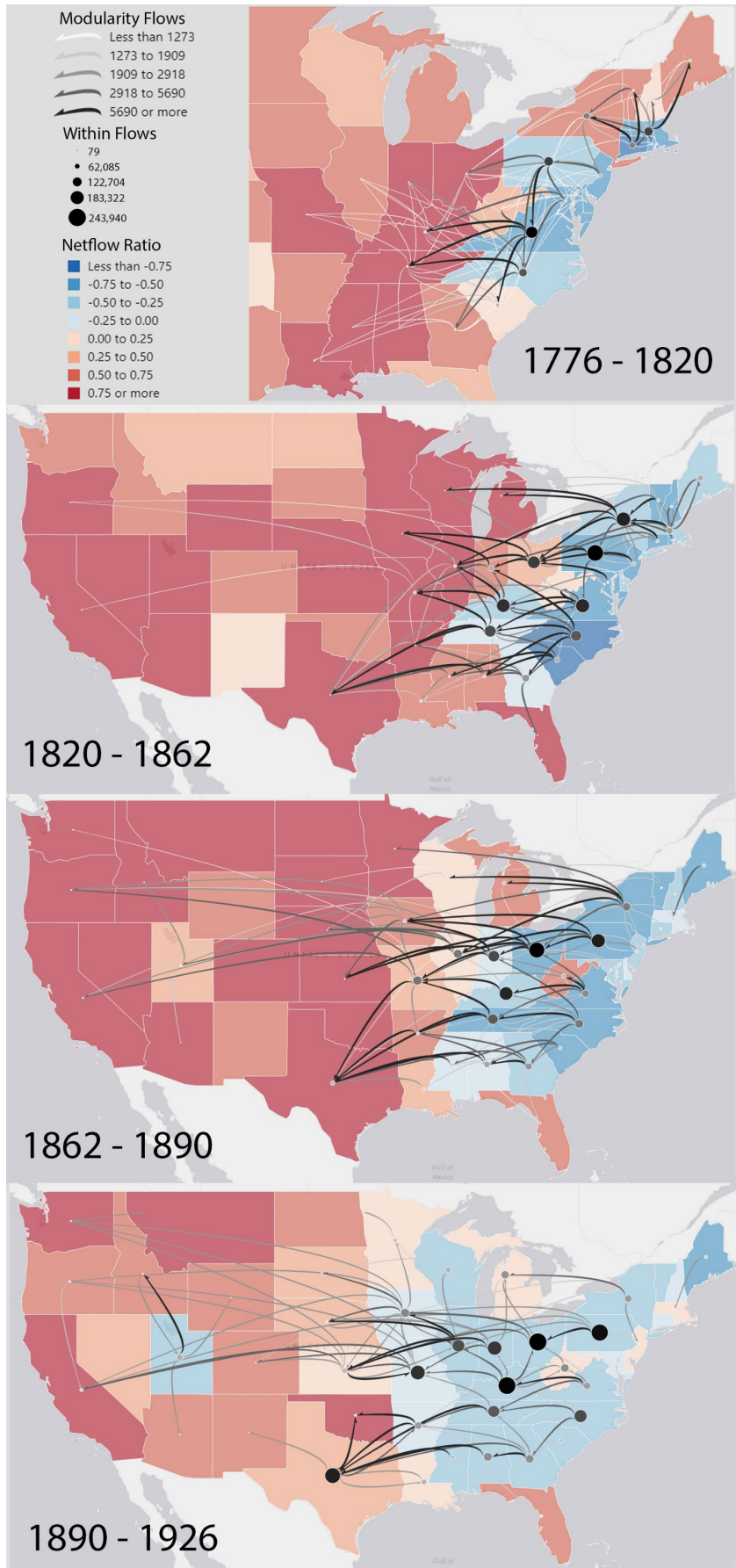

Figure 2: Migration of parent-child relations for the historical time periods.

**Acknowledgements:**

This paper is partly based upon work funded by Digging into Data Award LG-00-14-0030-14 by the Institute of Museum and Library Services (IMLS).